# CHIP-LEVEL CMP MODELING AND SMART DUMMY FOR HDP AND CONFORMAL CVD FILMS

George Y. Liu, Ray F. Zhang, and Kelvin Hsu
Intercon, 39120 Argonaut Way, Suite 320, Fremont, CA 94538
Tel: (510)-226-8730   Fax: (510)-490-3621

Lawrence Camilletti
Rockwell Semiconductor, Newport Beach, CA 92658

**ABSTRACT**

Chip-level CMP modeling is investigated to obtain the post-CMP film profile thickness across a die from its design layout file and a few film deposition and CMP parameters. The work covers both HDP and conformal CVD film. The experimental CMP results agree well with the modeled results.  Different algorithms for filling of dummy structure are compared.  A "smart" algorithm for dummy filling is presented, which achieves maximal pattern-density uniformity and CMP planarity.

## I. INTRODUCTION

Chemical Mechanical Polishing (CMP) has become an important semiconductor fabrication process in dielectric planarization, shallow trench isolation, and metal Damascene.  CMP process is also pattern sensitive although it provides the best planarity comparing to other techniques. The early modeling works on CMP pattern dependence are mainly at local level with some simple pattern structure.[1-5]  This is because chip-level CMP modeling requires a more sophisticated modeling algorithm and computing program for handling complex IC layout patterns. Local-level CMP modeling helps us to understand the basic nature of CMP pattern dependency. However, in order to be practically useful for the IC design and manufacturing, it would need chip-level modeling.  Recent effort in chip-level CMP modeling has showed some degree of success [6-7].  The chip-level modeling results can be useful for 1) acquiring planarity information before actual silicon wafer processing, 2) selectively filling dummies to widen the CMP process window and improve planarization, and 3) determine the thickest and thinnest dielectric positions on the die and obtain the intra-die ILD thickness variation range.  In this paper, we will present the results of chip-level CMP modeling for both HDP film and conformal CVD film.  Smart dummy filling based on the CMP modeling is will also be discussed.

## II. CHIP-LEVEL CMP MODELING AND EXPERIMENTAL RESULTS

CMP is well known to be sensitive to the pattern density.  However, the concept of pattern density is not numerically well defined, particularly for HDP film.  A clear definition of pattern density is essential to proceed a modeling.

For conformal CVD silicon oxide film, the side wall thickness is close to the deposition thickness.  When the spacing between the metal line is small, the gap is filled when dielectric film deposition is thick enough.  Therefore for fine pitch area, the elevated area density is significantly larger than the metal density as illustrated in Figure 1 (a).

Silicon dioxide film deposited by high-density plasma (HDP) technique has been widely employed for inter-layer dielectric (ILD) and shallow trench isolation (STI) because of its excellent capability in gap filling.  The topographic profile of HDP film exhibits a variety of complex patterns compared to the conformal CVD film, with the volumes of HDP film being dependent on the geometrical shape and size of the metal pattern (or active pattern in the case of STI) beneath, as illustrated in Figure 1 (b).

Hence, the pattern density of either HDP or conformal CVD film is very different from that of lithographic mask.  We define the pattern density as the ratio of unit area dielectric volume above he surface level of the open field to the metal thickness. For STI, metal thickness  should be replaced by trench depth. The definition given here is suitable for HDP, CVD, as well as composite film.

Figure 2 shows pattern density of HDP film and conformal CVD film deposited on top of metal of 50% pattern density at different pitch.  As it can be seen, the pattern density varies significantly with pitch size.

Chip-level CMP modeling is aimed to obtain the post-CMP film profile across a die from its design layout file and a few film deposition and CMP parameters.  To conduct the chip-level modeling, we first need to obtain chip-level pattern density map.  This requires to calculate volumes of various shapes of dielectric film on the top of all metal lines.  We accomplished this intensive computation from the GDS design layout file in a SUN workstation using CLIP (which is a commercial software package developed by Intercon).  Figure 3 shows the contour map of the pattern density of a test chip.  The test chip used in the study has a die size of 22×22 mm$^2$ and contains several SRAM and various test structures for process technology development.  The dielectric film consists of a HDP film with deposition-to-etch ratio of 4:1 followed by a conformal CVD oxide film.  A circular area with diameter of 2.5 mm is used to calculate the pattern density at the center of the circle.  The contribution of each point within the circle depends on the its distance to the center point and is weighted with a 2-D Gaussian function.

From the contour map in Figure 3, one can obtain the post-CMP dielectric thickness profile across the die.  As plotted in Figure 4, the modeled film thickness profiles along X and Y directions agree well with the results measured by a spectral photometer at fab. Figure 5 shown the histogram of the modeled thickness distribution across the whole die.

Charts like Figure 3 and Figure 5 are useful for CMP process characterization and optimization for a new IC product.  Positions with thickest and thinnest ILD can be easily determined from a contour plot like Figure 3. In the IC production line, post-CMP oxide thickness is often inspected at a specific metrology location in the scribe line area.  The ILD thickness at this metrology location generally dose not represent the thickness at the center of the distribution. The contour map and distribution plot can help an engineer to determine the optimal polishing thickness so that the center of the post-CMP thickness distribution is aligned to the specification target.

Figure 6 compares the pattern density distribution of the HDP film, CVD film, and design layout for a metal mask.  The HDP film has a smaller pattern density and narrower distribution than the design layout while conformal CVD film has a larger pattern density wider distribution than the design layout.  The wider density distribution generally results in a larger dielectric thickness variation.  Figure 6 explains why an HDP film generally provides better post-CMP planarity than a conformal CVD film.  Because HDP film has smaller pattern density, it requires less polishing time to planarize.  Consequently, the post-CMP dielectric thickness for HDP film have a better inter-die uniformity than that of conformal CVD film since the within-wafer-nonuniformity of CMP is generally much higher than that of film deposition.

## III. SMART DUMMY FILLING

The objective of dummy filling is to make pattern density more uniform across a die. Dummy metal has been proven to improve ILD planarization. Dummy active silicon has allowed a direct polish process for shallow trench isolation (STI) without costly etch-back process. In conventional methodology, dummies are filled to the open area with a dummy density fixed across the die. We will call the dummy filled under this methodology "conventional dummy". Conventional dummy generally does not guarantee a maximum uniformity of pattern density across the die. We have developed a computer software program that can locally select optimized dummy density to achieve maximum pattern-density uniformity across the die with minimum amount of dummies. We will call the dummy filled under this new methodology "smart dummy".

Figure 7 compares pattern density distributions for the cases without dummy, with conventional dummy, and with smart dummy. The STI mask of a product chip with size of $11\times14$ mm$^2$ was used for generating these plots. Trench depth of 4000 Å and conformal CVD oxide thickness of 6000 Å were assumed. Minimum distance between dummy and non-dummy feature was 3 μm.

The width of the pattern density distribution is proportional to the variation range of post-CMP intra-die dielectric thickness. Therefore, it should be maximally reduced in order to achieve minimum dielectric thickness variation. The plots in Figure 7 show that the smart dummy is more effective than the conventional dummy in reduction of the distribution width.

Dummy filling increases pattern density, which results in lower CMP throughput and higher inter-die oxide thickness variation. Dummy filling can also increase parasitic capacitance causing the reduction of device speed. Therefore, the usage of dummy should be minimized. To simultaneously achieve a narrow pattern density distribution and minimum increase of average pattern density, compromise must be made and a sophisticated computing algorithm is required. Figure 7 shows that the smart dummy cause less pattern density increase than the conventional dummy.

We note that the etch-back process for STI not only narrows pattern density distribution but also reduces average pattern density. To make direct polish STI process robust, dummy filling should also consider both the width of pattern density distribution and average pattern density.

## IV. CONCLUSION

The definition of pattern density applicable to both HDP film and conformal CVD film has been introduced. Chip-level CMP modeling based on the calculation of pattern density map and distribution has been presented. The modeled post-CMP dielectric thickness profile agrees well with the experimentally measured data. The modeling software should be useful for both IC layout design and CMP process development. A smart algorithm of dummy filling that locally selects optimal dummy density has been discussed. The algorithm minimizes the width of pattern density distribution with minimal amount of dummies. The post-CMP intra-die and inter-die dielectric thickness variation should be minimized when the algorithm is used for dummy filling.

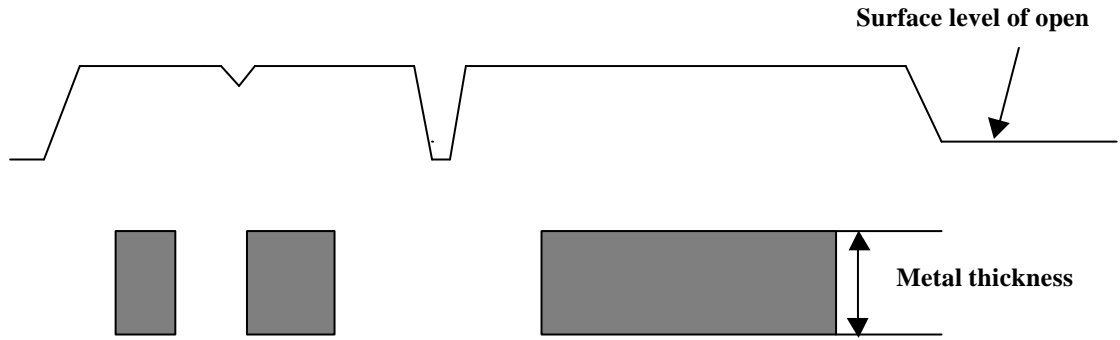

(a)

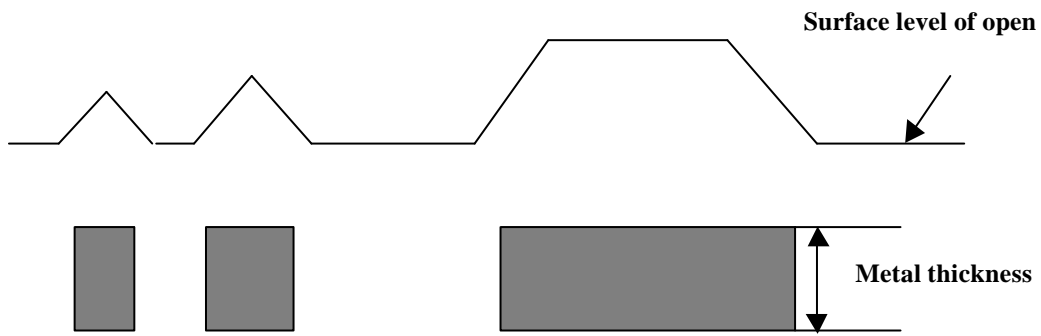

(b)

**Figure 1.** Illustration of surface profile of oxide film on the top of metal. (a) Conformal CVD oxide film. (b) HDP oxide film. The volume above the surface level of the open field is used for the pattern density calculation.

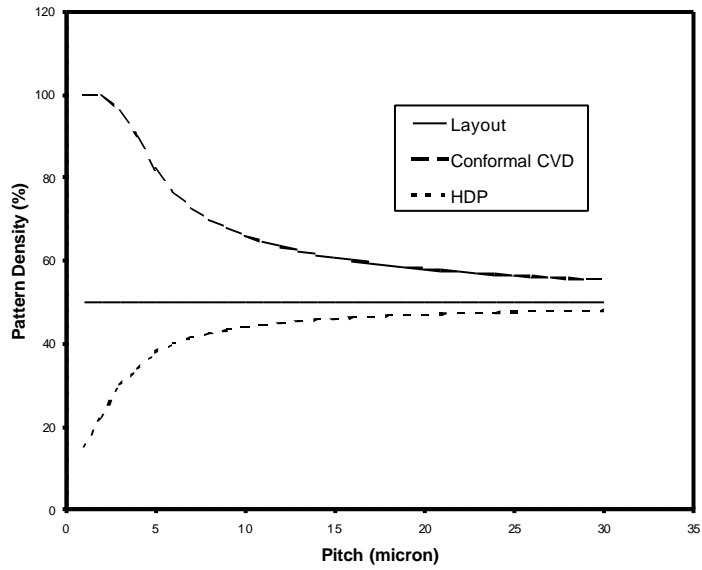

**Figure 2.** Comparison of pattern densities of HDP and conformal CVD film at different pitches for line arrays with the layout density of 50%.

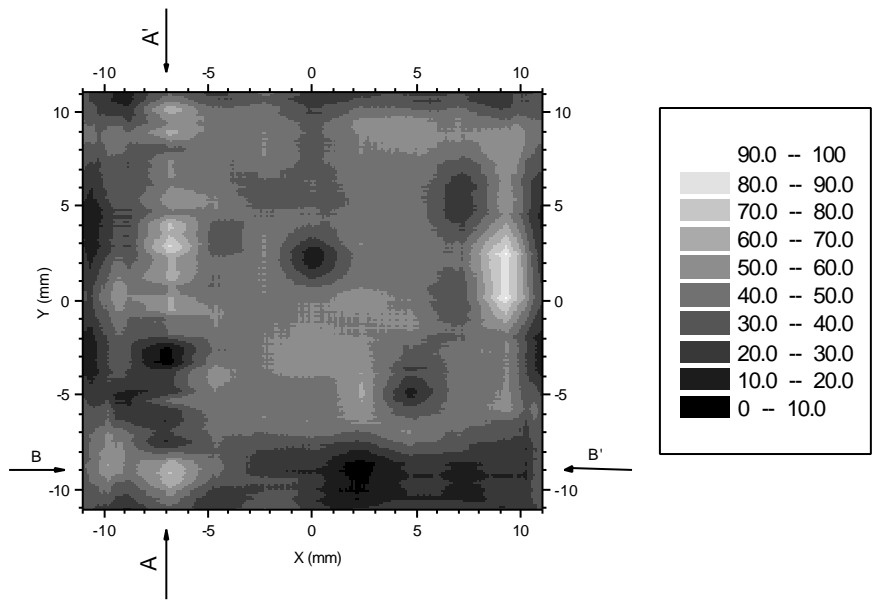

**Figure 3.** Pattern density map of a test chip for technology development. The locations of highest and lowest pattern density can be easily identified. Measurement data shown in Figure 4 (a) and (b) are taken along line AA' and BB'.

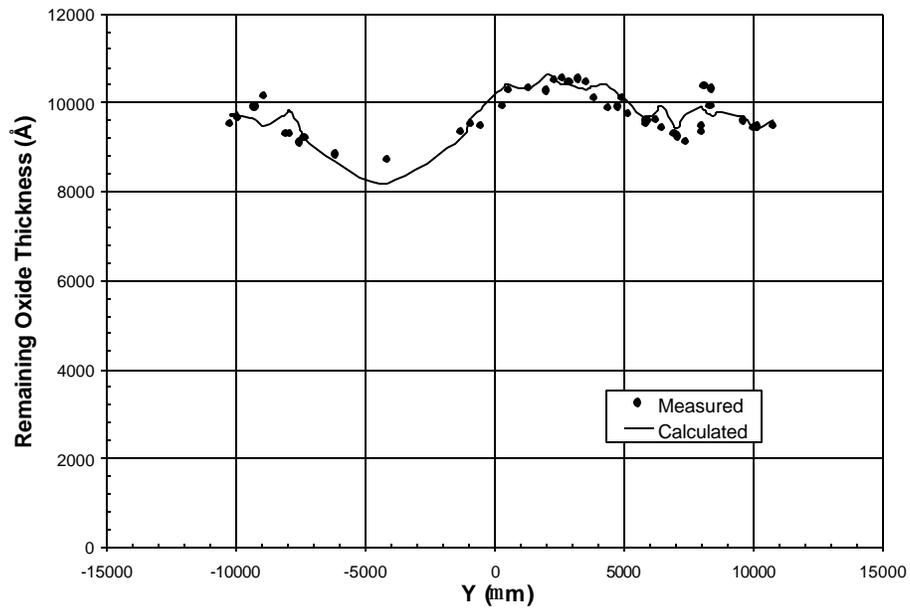

(a)

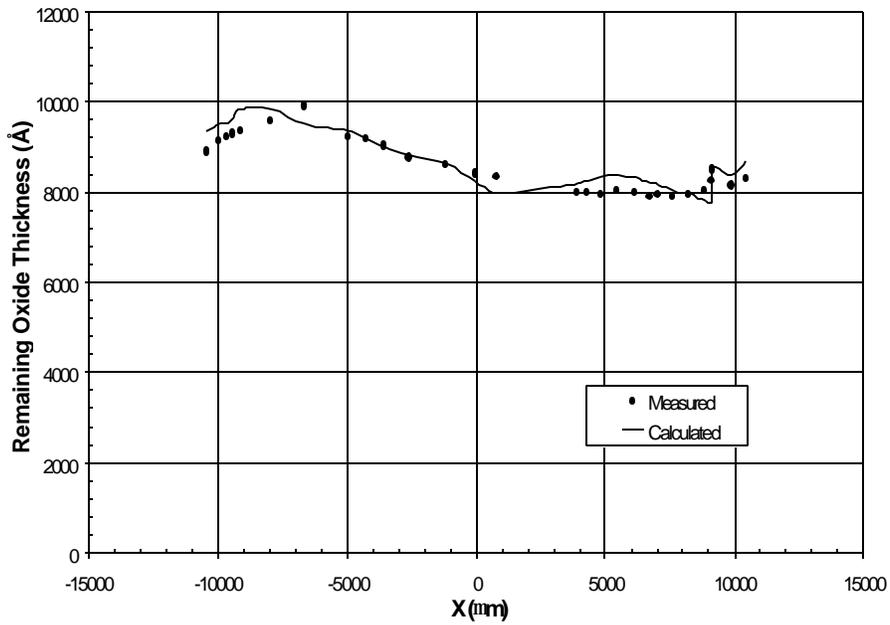

(b)

**Figure 4.** Comparison of measured and calculated oxide thickness. (a) along AA' line. (b) along BB' line. (see Figure 3 for illustration.)

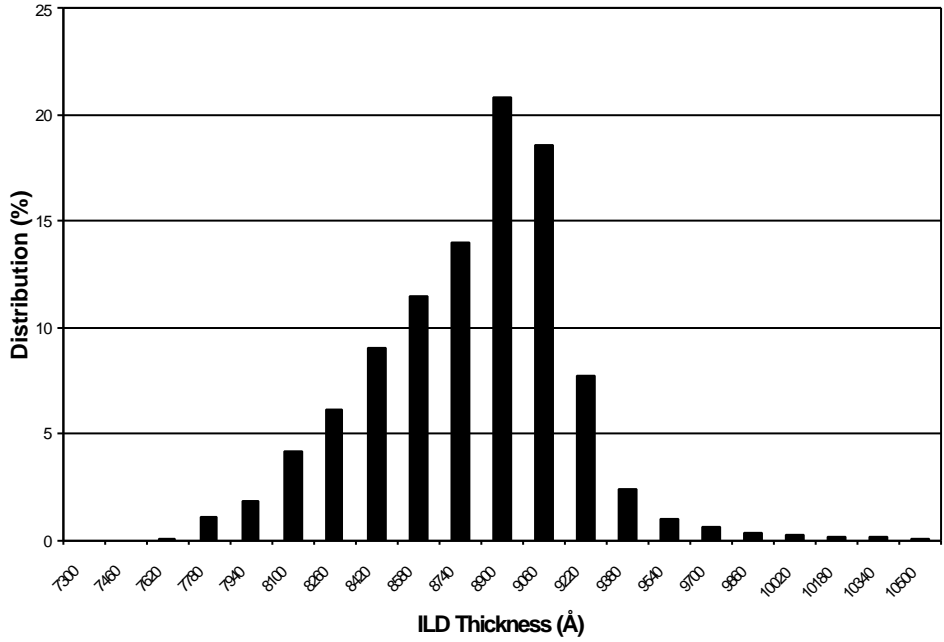

**Figure 5**. Modeled post-CMP ILD thickness distribution for the test chip.

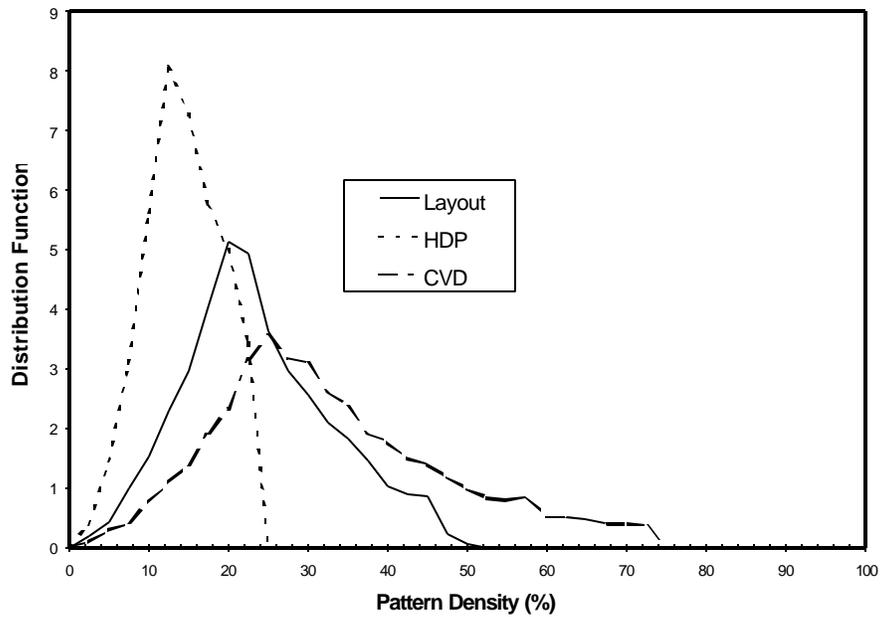

**Figure 6**. Comparison of pattern density of the HDP, comformal CVD, and design layout.

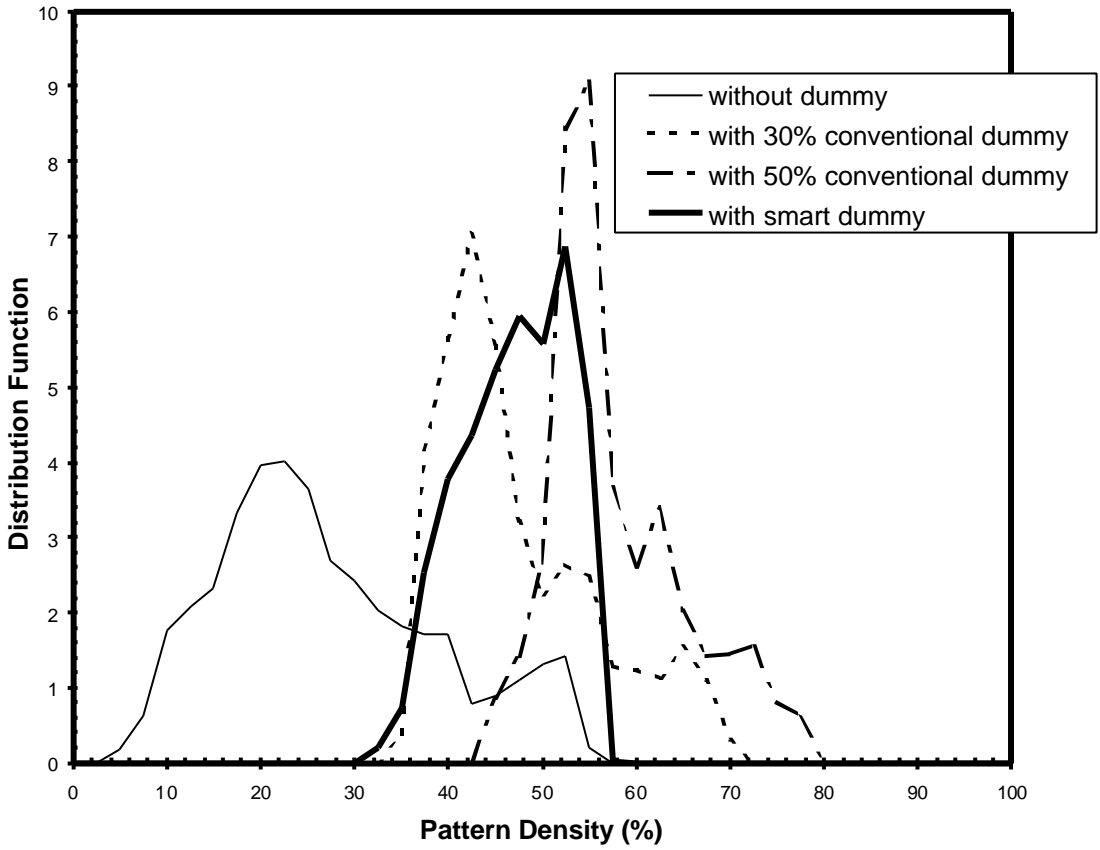

**Figure 7.** Comparison of pattern density distribution between conventional dummy filling and start dummy filling for a STI mask.